\documentclass[a4paper,11pt]{article}
\usepackage{pos}

\usepackage{epsf}

\def\NPB#1#2#3{{\it Nucl.\ Phys.}\/ {\bf B#1} (#2) #3}
\def\PLB#1#2#3{{\it Phys.\ Lett.}\/ {\bf B#1} (#2) #3}

\def\PRD#1#2#3{{\it Phys.\ Rev.}\/ {\bf D#1}  (#2) #3}
\def\PRL#1#2#3{{\it Phys.\ Rev.\ Lett.}\/ {\bf #1} (#2) #3}
\def\PRT#1#2#3{ {\it Phys.\ Rep.}\/ {\bf#1} (#2) #3}

\def\IJMP#1#2#3{ {\it Int.\ J.\ Mod.\ Phys.}\/ {\bf A#1} (#2) #3}

\def\APP#1#2#3{ {\it Astropart.\ Phys.}\/ {\bf #1} (#2) #3}
\def\EJP#1#2#3{ {\it Eur.\ Phys.\ Jour.}\/ {\bf C#1} (#2) #3}
\def\JHEP#1#2#3{ {\it JHEP}\/ {\bf #1} (#2) #3}

\def\etal{{\it et al\/}}

\def\AEF{A.E. Faraggi}

\newcommand{\beq}{\begin{equation}}
\newcommand{\eeq}{\end{equation}}
\newcommand{\beqa}{\begin{eqnarray}}
\newcommand{\beqn}{\begin{eqnarray}}
\newcommand{\eeqn}{\end{eqnarray}}
\newcommand{\eeqa}{\end{eqnarray}}

\newcommand{\ba}{\begin{eqnarray}}
\newcommand{\ea}{\end{eqnarray}}

\title{Experimental Particle Physics Priorities 2025:\\
\medskip
~~~~~~~~A String Phenomenology Perspective}

\author[c]{Jonathan DeMont}
\author*[a]{Alon E. Faraggi}
\author[b]{Mark Goodsell}
\author[c]{Marco Guzzi}

\affiliation[a]{Department of Mathematical Sciences, university of Liverpool, 
  Liverpool L69 7ZL, United Kingdom}

\affiliation[b]{Sorbonne Université, CNRS, Laboratoire de Physique Théorique et Hautes
Énergies (LPTHE), 75252 Paris, France}

\affiliation[c]{Department of Physics, Kennesaw State University, Kennesaw, GA 30144, USA}

\emailAdd{jdemont@students.kennesaw.edu}
\emailAdd{alon.faraggi@liv.ac.uk}
\emailAdd{mguzzi@kennesaw.edu}
\emailAdd{goodsell@lpthe.jussieu.fr}

\abstract{With the SNOWMASS 2021 process in the US and the on--going European Strategy Report 2025,
the field of elementary particle physics is undergoing detailed community evaluation, and
the experimental particle physics program, which requires substantial public
investment, is under scrutiny. We offer an assessment of the current experimental 
particle physics priorities from a string phenomenology point of view. String theory provides 
a perturbatively consistent framework for quantum gravity. String phenomenology aims to connect
between string theory and observational data. String theory is a consistent 
theory of quantum gravity that contains
the other fundamental constituents of matter and interactions. As all forms of energy couple 
to gravity, string theory provides a framework that reproduces the structures of 
the Standard Model of particle physics and gives rise to detailed physics 
scenarios beyond the Standard Model, {\it e.g.} dark matter candidates, axions, additional gauge
symmetries, etc. Given this breadth, we propose that from a string phenomenology 
perspective, the experimental particle physics priority is the nature of the Higgs boson and
the electroweak symmetry breaking mechanism. 
An ideal facility in the near future to study
this sector is a hadron collider at 50--60 TeV that utilises contemporary magnet 
technology and can be built in 10--15 years from decision. 
}

\FullConference{Proceedings of the Corfu Summer Institute 2024 "School and
Workshops on Elementary Particle Physics and Gravity" (CORFU2024)\
12 - 26 May, and 25 August - 27 September, 2024\\
Corfu, Greece\\}

\begin{document}
\maketitle

\section{Introduction}

With the discovery of the Higgs resonance by the ATLAS and CMS collaborations of the LHC experiment at CERN
\cite{ATLAS:2012yve, CMS:2012qbp},
the experimental validation of the Standard Model (SM) of particle physics
was completed. It is a pinnacle moment in the human journey of exploration, concluding a century of discoveries
that started with those of the electron and the nucleus by Thomson and Rutherford. 
The data that will continue to be accumulated by the LHC and the HL--LHC will further
improve the measurements of Standard Model parameters and possibly find deviations from it. 
The field of experimental particle physics is therefore at a crossroad. On the one hand, 
the Standard Model passes all experimental tests with flying colours. On the other hand, 
it still leaves many puzzles unanswered. In the first place, it is an ad--hoc collection of 
gauge symmetries and matter states, and of the parameters required to account for their masses
and interactions. While the number of parameters can be reduced by embedding the Standard Model 
in a Grand Unified Theory (GUT), the origin of the flavour replication and parameters remains a mystery. 
Neutrino masses vanish in the Standard Model and mandate its extension with more particles and parameters. 
The strong CP violating parameter is unnaturally small. Astrophysical and cosmological
observations mandate the existence of dark matter and dark energy, which the Standard Model
does not account for. Most perplexing of all is perhaps the nature of the Higgs particle and its
lightness relative to the Grand Unified Theory and the Planck scales. Naively, we expect the Higgs
particle mass to be at the cutoff scale, where the Standard Model no longer provides a viable
perturbative parameterisation of the observational data. But if the Standard Model remains
viable up to the GUT or Planck scales, then its mass should be at that scale, as it is not protected
by any symmetry. Possible solutions include the existence of a new fundamental symmetry, {\it e.g.}
supersymmetry, or that the Higgs is a composite, rather than a fundamental, particle. 
The Higgs sector of the Standard Model is the least studied and presents many questions
for future experiments. Is it fundamental or composite? Is there a single SM Higgs particle
or are there multi--Higgs particles like in the two Higgs doublet model? Is there 
new physics associated with electroweak symmetry breaking as anticipated due to the 
hierachy problem? The measurement of the Higgs cubic coupling and direct measurement 
of its coupling to the top quark are crucial for getting some insight into the Higgs 
potential and its stability. All these questions would have been probed by the defunct
Superconducting Super Collider (SSC) and must await the construction of Future Collider 
Facilities (FCF). 

All of the above puzzles warrant experimental exploration. All of them are driven by boundless human 
curiosity and deepest desire to address the most profound puzzles of our times. It is the age--old 
characteristic rooted in the basic craving of human beings. Understanding the world around us and 
how it came to be. There is, however, the question of priorities. Not only from a practical point 
of view, but also from the scientific perspective. Which is the experiment that is most likely to push the 
boundaries of human knowledge, as well as potentially break new ground with novel discoveries. 

We argue here that the experimental study of the Higgs sector of the Standard Model and beyond
is the most urgent experimental question of our time and should be given the most pressing priority.
The important points are the following. The Higgs particle was discovered at the LHC in 2012
but the parameters governing its dynamics, in particular its self--coupling are yet to be 
mesaured more precisely. Its properties will be studied at the High--luminosity LHC and will be 
further elucidated in future lepton and hadron colliders.  
Furthermore, while not guaranteed, it is in general anticipated that the Higgs particle 
of the Standard Model is augmented with additional particles and symmetries, such as 
the superparticle spectrum in the supersymmetric extension of the Standard Model, 
or with a new gauge sector which becomes strongly interacting not far above the 
electroweak scale and of which the Higgs particle is a composite state. Direct 
production of the new states, provided that they are within their energy 
reach, will ensue in future hadron colliders, whereas a lepton collider
may register their existence by observing deviations from precise measurements 
of the Higgs particle parameters. 

Another facet of the contemporary debate is that of technology development. Future
collider facilities will use novel ground-breaking technology. These new 
developments are exciting and interesting in their own right. One example of substantial
contemporary interest is that of a muon collider, where the development of the 
required technology is a fascinating enterprise. Another crucial technical aspect 
of new hadron colliders are the superconducting alloys that will be used in
the magnets and the stable magnetic field strength that they will be able to 
produce. Contemporary magnets at the LHC use Niobium–Titanium (NbTi) alloys. 
They operate at temperature of 1.9K and produce a stable magnetic field of 8.3 Tesla. 
Magnet design and manufacturing is
one of the key elements in both the success and the cost
of a collider experiment. 
Future hadron colliders
are planned to operate magnets with magnetic fields, 
of the order of 16 Tesla. 
The current alloy
technology cannot sustain stable magnetic fields of this
magnitude and new alloys need to be developed. The objective
is to construct superconducting magnets that are based
on the use of Niobium–Tin (Nb$_3$Sn) alloys to reach 
operating fields of 16 Tesla. However, the new alloy technology
is yet to be developed and as is always the case with new 
technology it may take more time than anticipated to develop it.
In addition to 
the scientific benefits, new accelerator physics developments 
can bring huge rewards in the commercial application 
of the new technologies in the wider economy. 

While new technology development is an exciting and well worth 
of pursuit and the allocation of resources, we advocate 
in this paper that priority should be given to the study
of the Higgs sector of the Standard Model and beyond using
existing LHC magnet technology that can produce stable magnetic
fields of the order of 10 Tesla. Using a tunnel similar to that 
envisioned in the Future Circular Collider, of the order of $\sim$91km, 
can yield a hadron collider with $\sim$50--60TeV CoM energy. 

Several clues point to the embedding of the Standard Model in
Grand Unified Theories (GUTs). The main evidence stems from the 
multiplet structure of its matter states. Each of the chiral 
generations, including the right--handed neutrino
fits in a spinorial representation of $SO(10)$, thus
reducing the arbitrariness in their charge assignments. 
Additional evidence stems from the logarithmic running 
of the Standard Model gauge parameters, which notably in supersymmetry is compatible 
with the unification hypothesis, as well as the successful 
running of the mass parameters in the heavy generation
but not in the lighter two generations. 
The suppression of left--handed neutrino masses
is explained neatly by the  high scale seesaw mechanism.
Proton longevity similarly favours unification at a high scale. 
In the Standard Model, lepton and baryon numbers are accidental global
symmetries at the renormalisable level, and we anticipate 
in general that extensions of the Standard Model will produce proton-decay mediating operators, suppressed by some cutoff scale. 
The longevity of the proton suggests that the cutoff scale 
has to be of the order of the GUT scale or above. 

\begin{figure}
    \centering
    \includegraphics[width=0.7\linewidth]{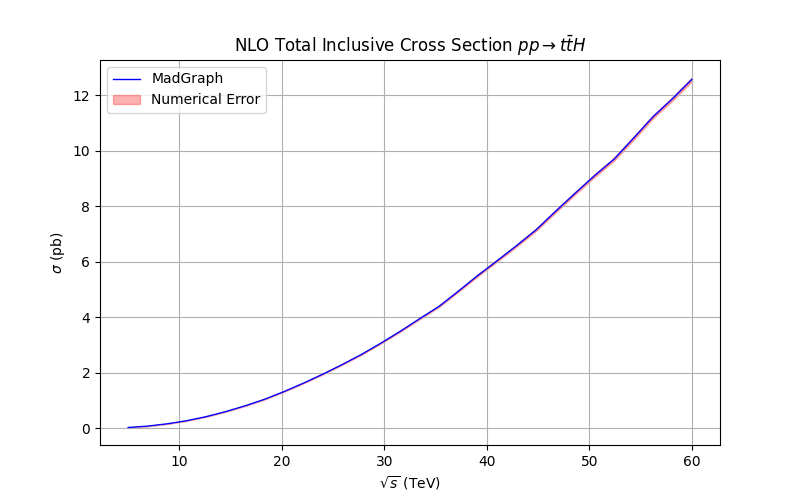}
    \caption{Total inclusive cross section for Higgs boson production in
    association with a top-antitop quark pair in proton-proton collisions
    as a function of the collision energy. The cross section calculation is
    obtained with Madgraph~\cite{Alwall:2014hca} at next-to-leading order 
    (NLO) in QCD using the CT18NLO PDFs~\cite{Hou:2019efy}.}
    \label{fig1}
\end{figure}

These properties of the Standard Model provide very compelling 
motivation for its embedding in $SO(10)$ GUT. However, this 
embedding still does not provide an explanation for the flavour structure
of the Standard Model. The fundamental origin of the flavour 
parameters can only be sought by fusing the Standard Model
with gravity. String theory provides the arena to explore the synthesis 
of the Standard Model with gravity in a self-consistent
framework. The self-consistency conditions of string theory
produce an extended flavour structure which is related to the 
properties of the internal compactified manifolds of string 
vacua. Indeed, detailed flavour parameters can be calculated 
and reproduce the gross flavour structure of the Standard Model. 
However, there is a crucial point that cannot be over-emphasised. 
In all these examples the Higgs state that reproduces the flavour 
structure of the Standard Model is a fundamental scalar state
that couples to the fundamental fermion matter states of the Standard 
Model via the contact terms of their scattering amplitudes. 
Thus, if it turns out that the Higgs resonance observed at the LHC and
future collider facilities is a composite particle, rather than a 
fundamental scalar state, then a whole slew of phenomenological 
string vacua may be cast away as they will not be relevant 
to the observational data. 
As of today, there is not a single composite Higgs model 
from string theory, let alone one that can reproduce the gross
flavour structure of the Standard Model. This is not to say that one may not exist, 
but it will require a complete paradigm shift in attempts to 
construct quasi--realistic string models. Thus, from a string 
phenomenology perspective, the nature of the Higgs boson
is of primary importance. In this respect, it is anticipated 
that forthcoming experimental facilities that are currently under
consideration, will only be able to set limits on the composite scale
of the Higgs boson. That is an acceptable risk that we have to take.
The collider facilities will be able to "push the envelope" on that question. 
They will improve the measurements of the Standard Model electroweak symmetry breaking sector.
In particular, they will improve the precision of the cross-section measurement for 
reactions whose cross-section is small as measured at current facilities such as the LHC (see Fig.~\ref{fig1}).
This will pave the way to a new realm of precision, further expanding the boundary of
precision calculations in Quantum Field Theory.   
They will potentially discover new physics 
associated with the electroweak symmetry breaking mechanism. Most importantly, 
these questions are well defined from an experimental point of view. 
The Higgs boson exists and is waiting to be studied and explored. 
One could not be more excited with curiosity and anticipation. 

The self-consistency conditions of string theory
imply the existence of new sectors, {\it e.g.} the hidden sector in the 
heterotic-string, and gives rise to possible signatures beyond the Standard 
Model, like dark matter candidates; axion and axion-like particles; 
new Abelian gauge symmetries; and more. 
Many of these extensions of the Standard Model are considered in the 
Beyond the Standard Model (BSM) phenomenology, using point--particle 
quantum field theory methods. 
While the string phenomenology scenarios are constrained 
by the self-consistency conditions of string theory, the field 
theory BSM scenarios are not similarly constrained and substantial 
more freedom is afforded. An example, is the case of extra Abelian
vector-bosons in BSM models that have been studied since the 
mid-eighties \cite{zpreports}, whereas a string derived $Z^\prime$ model that can 
remain unbroken down to low scales was by far more difficult to obtain \cite{frzprime}. 

In the following, we will describe the framework of phenomenological 
string models in the free fermionic formulation that can serve as benchmark
models, the flavour structure they can give rise to, and possible signatures 
beyond the Standard Model. We will consider these extensions in relation to 
experimental searches that may look for them. Our aim is to argue that while 
these are compelling examples for physics beyond the Standard Model, 
they do not present a well defined proposition from an experimental point of 
view, and therefore should be given lower priority compared to the experimental
study of the Higgs boson. 

\section{Fermionic $Z_2\times Z_2$ orbifolds}

\begin{figure}[h]
\includegraphics[width=20pc]{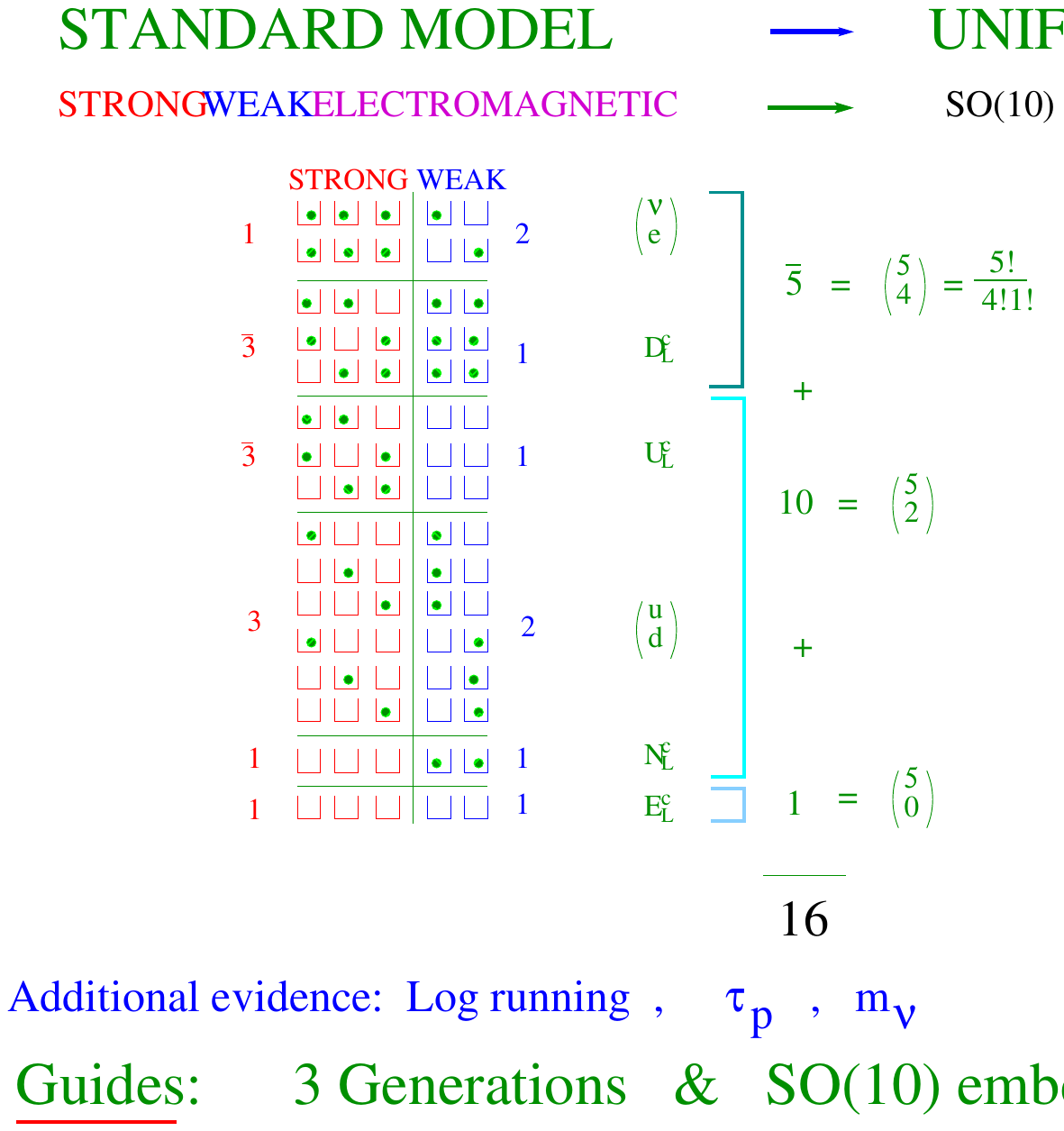}\hspace{2pc}%
\begin{minipage}[b]{10pc}
\caption{\label{figure1}%
\emph{
The Standard Model (SM) matter states strongly favours
their embedding in Grand Unified Theory (GUT) structures. 
Particularly compelling is that of $SO(10)$ in which 
each Standard Model family fits into a single spinorial
$16$ representation of $SO(10)$.  Additional 
evidence for high--scale unification stems from: logarithmic 
running of the SM parameters; proton lifetime;
suppression of left--handed neutrino masses.
}
}
\end{minipage}
\end{figure}

The Standard Model (SM) of particle physics, augmented with a right-handed neutrino to 
allow for neutrino masses, accounts for all sub-atomic observational data to date.  
The true essence of the Standard Model is revealed by the form of its 
multiplet structure, as shown in figure \ref{figure1}. The SM strongly favours
the realisation of Grant Unification structures in nature. 
Particularly compelling is its embedding in $SO(10)$ Grand Unified Theory
in which each one of the SM families is embedded in a single spinorial $16$
representation of $SO(10)$. If we count the SM gauge charges as experimental 
observable, the embedding in $SO(10)$ reduces the number of 
parameters needed to account for the SM matter gauge charges from 54 
(3 generations times 3 group factors times 6 multiplets) to 1, which is the
number of spinorial $16$ representations of $SO(10)$ needed to 
account for the SM matter states. A remarkable coincidence indeed. 
Additional
evidence for the realisation of GUT structures in nature is provided by
the logarithmic running of the Standard Model parameters; 
by proton longevity; and by the suppression of left--handed neutrino masses.  

\begin{figure}[h]
\includegraphics[width=20pc]{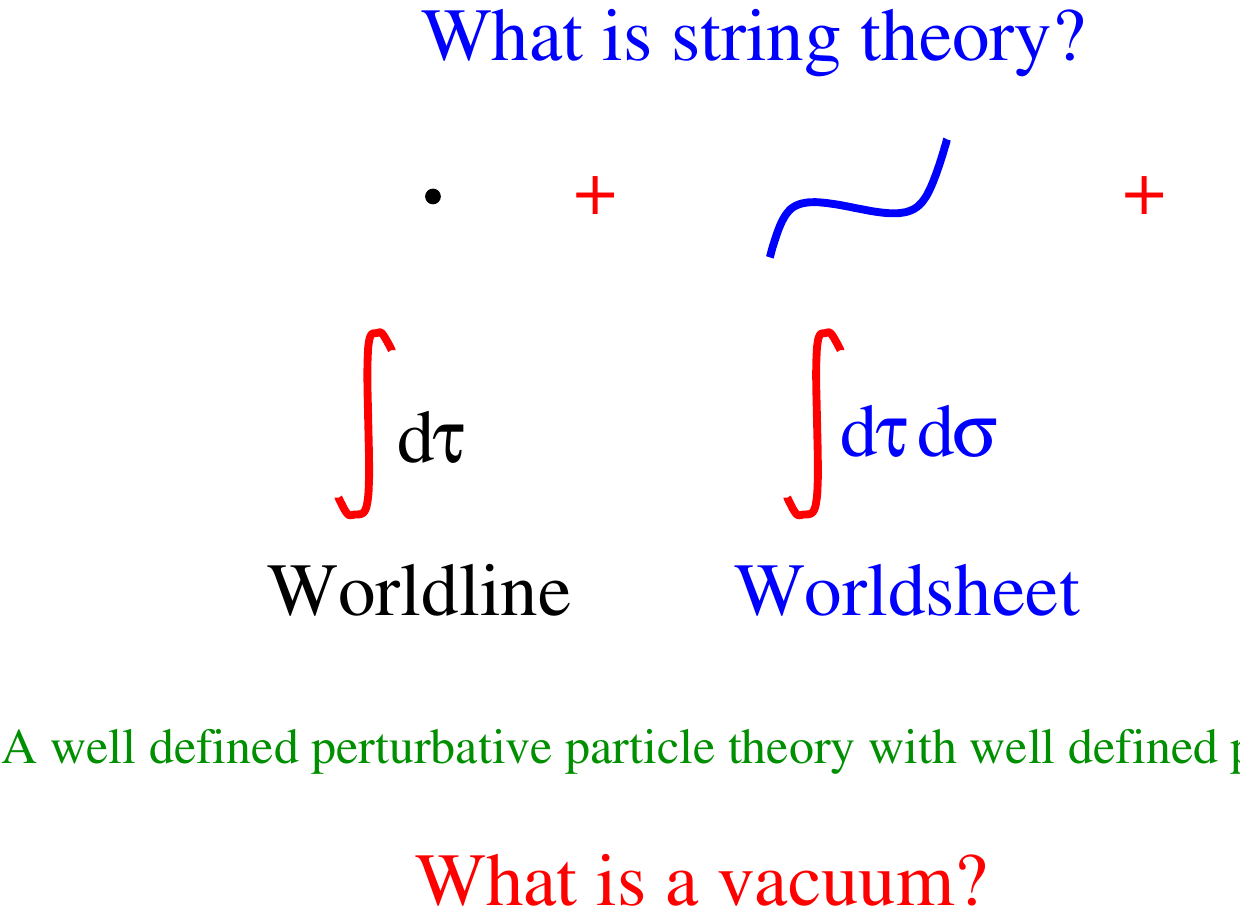}\hspace{2pc}%
\begin{minipage}[b]{10pc}
\caption{\label{whatisstringtheory}%
\emph{
What is string theory? A well defined perturbative particle theory
with well defined physical attributes. 
}
}
\end{minipage}
\end{figure}

The compelling picture of Grand Unified Theories is, however, not complete.
In particular, the replication of the matter generations and the mass and mixing 
data remains ad hoc. To find a fundamental origin for these parameters necessitates
the synthesis of the quantum gauge field theories with gravity. The most developed 
contemporary framework to explore these questions is that of string theory. String theory is
an extension of point quantum field theories, in which we parametrise the 
paths of elementary paticles
using one worldline parameter. In the string theory on the other
hand we parametrise the paths of elementary particles using two worldsheet parameters.
This is dictated for the consistency of the formalism with quantum gravity. Both 
point-like quantum field theories and string theories provide a well defined
perturbative particle theories with well defined physical attributes, in the form
of spacetime and internal charges. In both cases, what enables having well defined
physical attributes is the existence of a well defined vacuum. This is highly 
non-trivial and in this respect we should replace the often discussed question
of "What is string theory?" with "What is a vacuum?". Figure \ref{whatisstringtheory}
provides a graphic depiction of this argument. 

\medskip

\begin{figure}[h]
\includegraphics[width=16pc]{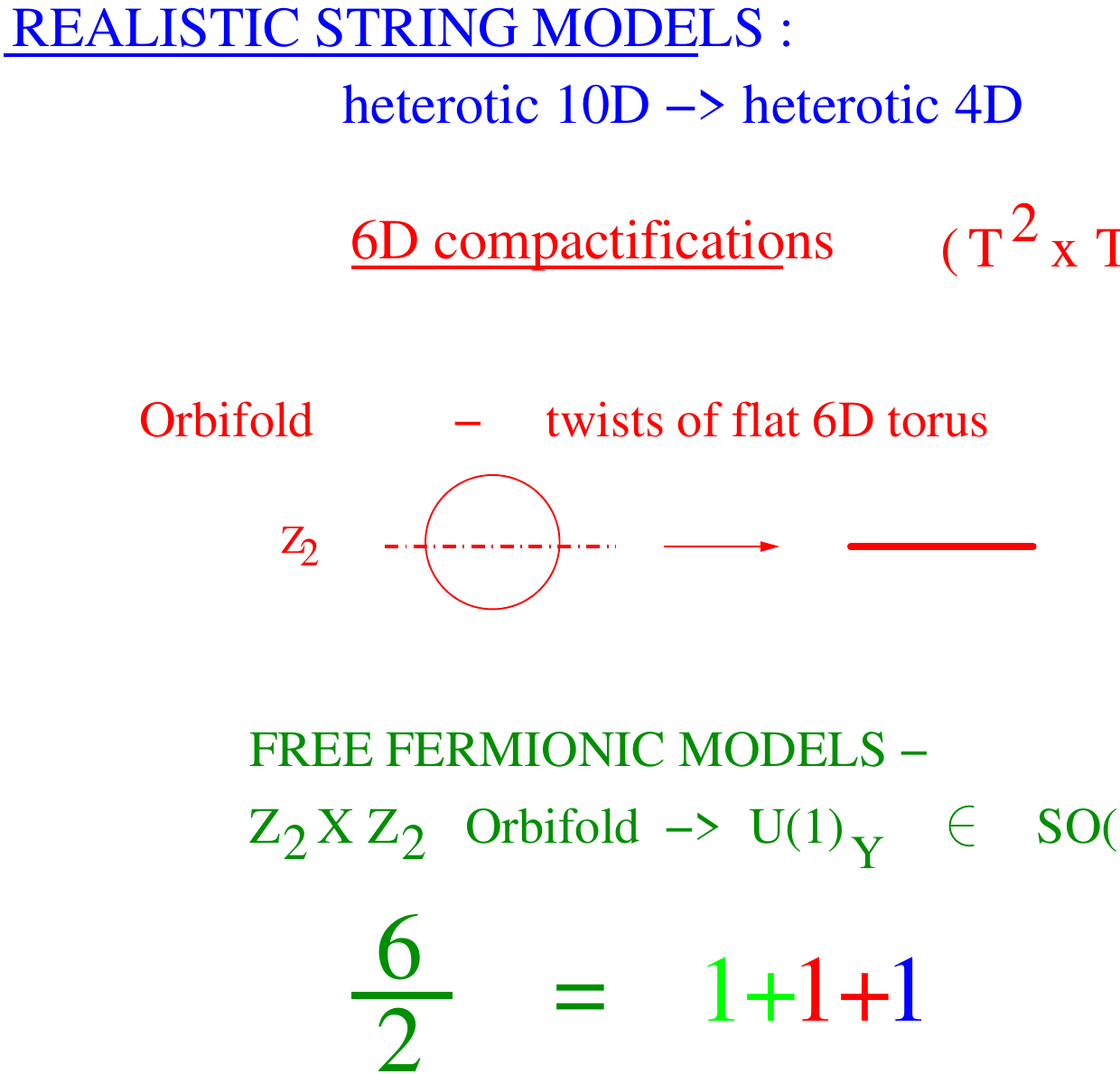}\hspace{2pc}%
\begin{minipage}[b]{10pc}
\caption{\label{RSM}%
\emph{
The fermionic $Z_2\times Z_2$ orbifolds produce an abundance of
three generation models with the canonical GUT embedding of the
weak hypercharge $U(1)_Y$. 
}
}
\end{minipage}
\end{figure}

We discussed above the $SO(10)$ GUT structure which is motivated by the Standard Model
data. However, building $SO(10)$ GUT field theory models presents numerous challenges,
like doublet-triplet splitting and $SO(10)$ gauge symmetry breaking, 
which requires utilisation of large representations, and make the field theory 
analysis quite cumbersome. Obtaining the underlying $SO(10)$ GUT structure 
from string theory gives rise to a stringy doublet-triplet splitting mechanism \cite{ps}
in which the harmful triplets are projected out of the massless spectrum 
by the Generalised GSO projections and the electroweak doublets remain in the 
spectrum, and the $SO(10)$ GUT symmetry is broken directly at the string scale 
by Wilson-line breaking. The construction of phenomenological GUT string models 
proceeds by compactifying the $E_8\times E_8$ heterotic-string on a
six-dimensional internal manifolds. The simplest constructions
are compactifications on six-dimensional flat torii, that are often taken
as a product of three $T^2$ torii. The torii are modded out by some internal symmetries
which gives rise to orbifold fixed points. The simplest example is that of $S^1/Z_2$
in which a circle is modded out by the reflection symmetry across the real axis, 
which produces a segment with two orbifold fixed points at the end points of the 
segment. Generalising to the $T^6/(Z_2\times Z_2)$ produces the $Z_2\times Z_2$
orbifolds of the heterotic-string in four dimensions. This class of string
compactifications have been mostly studied by using the free fermionic formulation
\cite{fff},
which provides a robust algorithmic formalism to analyse the massless string spectrum
and interactions, as well as the partition functions and the massive 
string spectrum. One important feature of the $T^6/(Z_2\times Z_2)$
orbifold is that it gives rise to three twisted sectors, and we may attribute the
origin of three generations in nature to the existence of these three twisted sectors, 
where each generation is obtained from a separate twisted sector \cite{nahe}. 
Figure \ref{RSM} depicts qualitatively the construction of fermionic $Z_2\times Z_2$
orbifold and their robust phenomenological properties being the existence of three 
generations and the canonical GUT embedding of the weak hypercharge $U(1)_Y$. 
\medskip
It is important to 
note that if the basic criteria that are taken as a guide
in the construction of the phenomenological string vacua are the existence of 
three generations and their $SO(10)$ embedding, then the perturbative string theory that
should be used is the heterotic $E_8\times E_8$ string because it is the only one that 
gives rise to spinorial 16 representations of $SO(10)$ in the massless spectrum. The
heterotic $E_8\times E_8$ string is conjectured to be
an effective limit of a more fundamental theory
about which we know virtually nothing.

String theory provides the tools to develop a phenomenological approach to quantum gravity.
Three generation quasi-realistic models that admit the $SO(10)$ embedding of
the Standard Model matter states were constructed in the Free Fermionic Formulation (FFF) 
\cite{fff} of the heterotic-string in four dimensions. 
These models provide benchmark models to explore how the parameters of the 
Standard Model may be determined from a theory of quantum gravity.
Many of the questions pertaining to the phenomenology of the Standard 
Model and Grand Unified Theories were studied in these models. Among them: 
$\underline{\hbox{Top quark mass prediction}}$,
at a mass scale of $O(175-180)$GeV \cite{tqmp}, which was predicted
four years prior to its experimental observation \cite{topdiscovery}; 
textures of the Standard Model charged leptons and quark 
mass and mixing matrices \cite{fermionmasses}, as well as
left-handed neutrino masses \cite{nmasses}; 
proton stability \cite{ps}; 
string gauge coupling unification \cite{gcu};
squark degeneracy \cite{sd}; 
and
moduli fixing \cite{moduli}. 
Moreover, the free fermionic models produced the first examples
of string models that give rise solely to the spectrum of the 
Minimal Supersymmetric Standard Model (MSSM) in the low energy effective field 
theory of the Standard Model charged sector.  We refer to such models as 
Minimal Standard Heterotic String Models \cite{mshsm}.

All the extra degrees of freedom needed to cancel the 
conformal anomaly are represented 
in the fermionic construction of the heterotic-string
in four dimensions in terms of free fermions propagating on the 
string worldsheet \cite{fff}. In the standard notation
the sixty-four worldsheet fermions in the lightcone gauge are denoted by: 

\leftline{~~~
${\underline{{\hbox{Left-Movers}}}}$:%
~~~~$\psi^\mu_{1,2},~~~~{ \chi_i},~~~~{ y_i,~~\omega_i}~~~~{(i=1,\cdots,6)}$}
\vspace{4mm}
{\leftline{~~~${\underline{{\hbox{Right-Movers}}}}$:}}%
$$~~~{\bar\phi}_{A=1,\cdots,44}~~=~~
\begin{cases}
~~{ {\bar y}_i~,~ {\bar\omega}_i} & ~~~{ i=1,{\cdots},6}\cr%
  & \cr%
~~{ {\bar\eta}_i} & ~~~{ U(1)_i ~~~i=1,2,3}~~\cr%
~~{ {\bar\psi}_{1,\cdots,5}} & ~~~SO(10) \cr%
~~{{\bar\phi}_{1,\cdots,8}}  & ~~~{ SO(16)}
\end{cases}\label{worldsheetfermions}
$$%
where the six compactified toroidal coordinates correspond to 
$\{y,\omega\vert{\bar y},{\bar\omega}\}^{1,\cdots,6}$ and the gauge group
symmetries are produced by the sixteen complexified right-moving fermions. 
String models in the free fermionic formulation are constructed in terms of 
a set of boundary condition basis vectors 
and the Generalised GSO (GGSO) projection coefficients of the one loop 
partition function \cite{fff}. The free fermion models correspond to 
$Z_2\times Z_2$ orbifolds with discrete Wilson lines \cite{z2xz2}. 

Quasi-realistic three generations models were 
constructed in the FFF since the late eighties \cite{mshsm, fsu5, so64, lrs}. 
The early models were highlighted examples that shared a common GUT structure 
generated by the so--called NAHE set \cite{nahe}, which consists of five 
basis vectors denoted as $\{ {\bf1}, S, b_1, b_2, b_3\}$. The gauge symmetry at this
level is $SO(10)\times SO(6)^3\times E_8$, with forty--eight
generations in the spinorial {\bf 16} representation of $SO(10)$, obtained
from the twisted sectors of the $Z_2\times Z_2$ orbifold $b_1$, $b_2$
and $b_3$. The $S$--vector produces $N=4$ spacetime supersymmetry, 
which is broken to $N=2$ by the basis vector $b_1$ and to $N=1$ by the 
inclusion of both $b_1$ and $b_2$.  
The second stage in the NAHE--based free fermionic heterotic--string
model building consists of adding three additional basis vectors
to the NAHE--set that break the $SO(10)$ gauge symmetry to one of its 
subgroups and reduce the number of generations to three. 
Each of the sectors $b_1$, $b_2$ and $b_3$ gives rise to one generation
that form complete {\bf 16} multiplet of $SO(10)$. The models 
admit the needed scalar states to further reduce the gauge symmetry and
to produce a viable fermion mass and mixing spectrum 
\cite{tqmp,fermionmasses,nmasses}.

For our purpose here, the old-school construction of three generation NAHE-based 
models suffices to illustrate how these string models gives rise to quasi-realistic
fermion masses and mixings, which is obtained due to the coupling of the 
matter spectrum from the sectors $b_1$, $b_2$ and $b_2$, to the fundamental Higgs states 
in the models that are obtained typically from the Neveu-Schwarz sector, and 
the sector $b_1+b_2+\alpha+\beta$. These sectors produce massless states
in the vectorial 10 representation of $SO(10)$, which are reduced to the
components of the unbroken $SO(10)$ subgroup. We note in passing that over the past 20+
years, systematic classification methods of fermionic $Z_2\times Z_2$ orbifolds 
were developed 
\cite{so10class, so64class, fsu5class, slmclass, LRSclass, fmp} 
, which enabled the scanning of a large number of string vacua and
the extraction of some of their general properties \cite{svd}. 

\section{Fermion masses and mixing from string theory}

Perhaps the main interest in phenomenological string constructions is the fact that they enable the calculation of the SM flavour parameters. The free fermionic standard-like models 
\cite{mshsm} illustrate this point. These models generically give rise to three generations from the twisted sectors $b_1$, $b_2$ and $b_3$. Three pairs of Electroweak Higgs doublets, 
$h_i,~{\bar h}_i, i=1,2,3$ arise from the untwisted sector. Additional one or two pairs, 
$h_{\alpha\beta},~{\bar h}_{\alpha\beta}$, are obtained from the sector
$b_1+b_2+\alpha+\beta$. The models typically contain additional states that
are $SO(10)$ singlets and exotic states that carry fractional charges with respect to 
either the weak hypercharge or the extra $U(1)_{Z^\prime}$ symmetry in $SO(10)$.
The exotic states are vector-like and typically get heavy mass along flat directions of 
the superpotential \cite{fc}, and in some models do not arise as massless states
\cite{so64class}. The string models contain $SO(10)$ singlet fields that may obtain
non-trivial Vacuum-Expectation-Values (VEVs) along supersymmetric flat directions. 
Fermion and scalar mass terms in the superpotential are of the form
\begin{equation}
    {cg}{f_if_j}h{{\left({{\langle\phi\rangle}\over{M}}\right)^{N-3}}}
\end{equation}
where $c$ are calculable coefficients from the tree level correlators between vertex oparators; $g$ is the gauge coupling which is fixed by the dilaton VEV; $f_i$ and $f_j$ are
the fermionic states from the sectors $b_{1,2,3}$; $h$ stands schematically for the 
light Higgs multiplets in the effective field theory limit; $M~\sim~10^{18}~GeV$
is related to the string scale; and
$\langle\phi\rangle$ are VEVs of Standard Model singlet fields along supersymmetric
flat directions. These VEVs are typically of the order of $0.1 M$ generating fermion mass terms
in the low energy effective field theory that are suppressed relative to the leading 
top-quark mass term \cite{tqmp}. The calculation 
of Standard Model fermion masses yielded a viable prediction for
top quark mass prior to its experimental observation \cite{tqmp}.
The top quark Yukawa coupling is obtained at the cubic level of the superpotential,
yielding $\lambda_t=\langle Q t_L^c H\rangle = \sqrt2 g$.
The Yukawa couplings for the tau lepton and bottom quark 
are obtained from quartic order terms. The quartic order
coefficients are calculated using Conformal Field Theory methods, and the 
VEVs of the SM singlet field is extracted from analysis of the
supersymmetric flat directions.
This calculation gives effective Yukawa couplings for the 
tau lepton and bottom quark in terms of the unified gauge coupling 
given by $\lambda_b=\lambda_\tau=0.35 g^3\sim~{1/8}\lambda_t$ \cite{tqmp}.
The Yukawa couplings at the string 
scale are extrapolated to the electroweak scale using the
Minimal Supersymmetric Standard Model (MSSM) Renormalisation
Group Equations (RGEs). The unified 
gauge coupling at the string scale is assumed to be compatible with the value 
required by the gauge coupling data at the electroweak scale. 
The bottom Yukawa is further extrapolated to the
bottom mass scale, which is used to obtain a value for $\tan\beta=v_1/v_2$,
where $v_1$ and $v_2$ are the VEVs of the two MSSM electroweak Higgs 
doublets. The top quark mass is then given by
$$m_t~=~\lambda_t(m_t)~{{{v}_0}\over{\sqrt2}}~{\tan\beta\over
{(1+\tan^2\beta)^{1\over2}}}$$
with $v_0=\sqrt{2(v_1^2+v_2^2)}=246{\rm GeV}$, yielding
$m_t\sim175-180{\rm GeV}$ \cite{tqmp}.

The analysis of the Yukawa couplings for the lighter
two generations progresses by analysing higher order terms 
in the superpotential and extracting the effective
dimension four operators \cite{superpoterm}.
The explorations to date included, for example, 
demonstration of the generation mass hierarchy and
Cabibbo--Kobayashi--Maskawa (CKM) mixing \cite{fermionmasses};
neutrino masses \cite{nmasses}. An illustration of this 
analysis \cite {fermionmasses} is provided in eq. (\ref{cabibbomixing}) 
\begin{equation}
    M_d\sim\left(
\begin{matrix}
                             \epsilon &{{V_2{\bar V}_3\Phi_{\alpha\beta}}\over{M^3}} &0 \cr
 {{V_2{\bar V}_3\Phi_{\alpha\beta}\xi_1}\over{M^4}} &{{{\bar\Phi}_2^-\xi_1}\over{M^2}} &0 \cr
                                               0 &        0 &{{\Phi_1^+\xi_2}\over{M^2}}\cr
\end{matrix}
\right)v_2,
\label{cabibbomixing}
\end{equation}
with 
$$\epsilon<10^{-8},~ 
\frac{V_2{\bar V}_3\Phi_{\alpha\beta}}{M^3}~=~\frac{\sqrt{5}~g^6}{64~\pi^3}
~\approx ~2-3\times 10^{-4},$$
Diagolnolising $M_D$ by a bi--unitary transformation yields the 
Cabibbo mixing matrix \cite{fermionmasses}
$$~~\vert V\vert~\sim~
\left(
\begin{matrix}
0.98&0.2&0\cr 
 0.2&0.98&0\cr 
 0&0&1\cr
\end{matrix}
\right),$$
whereas extended solutions to three generation mixing are considered in 
\cite{fermionmasses}.
These examples illustrates that the string constructions provide the arena 
to calculate the basic mass and mixing parameters in the Standard Model. 
This statement should be taken with caution and there are many issues
and problems that need to be addressed before a fully reliable calculational
framework is established. However, the crucial property of the string
models that facilitate the development of this framework is
the existence of fundamental scalar electroweak doublet representations
and their couplings to the fermion matter states in the string spectrum.
From this point of view the crucial question to forthcoming
particle physics experiments is: 

\begin{center}
    {\bf is the Higgs particle fundamental or composite?}, 
\end{center}

\noindent
which is a question that forthcoming Collider Facilities
(FCF) will be able to study and address, or in the very least, set limits
on the scale of compositeness of the Higgs boson. It is vital to stress that
from an experimental point of view this is a well defined enterprise, with
significant benefit to the advance of our basic understanding of nature, 
and significant potential for prizes and accolade in the process of
experimental exploration and discovery. 

\subsection{Probing compositeness}

Composite Higgs models (see e.g. \cite{Contino:2010rs} for a review), invoked to solve the hierarchy problem,  are considered natural provided that the hyper-pion decay constant $f_\Pi$ is below $\mathcal{O}(\mathrm{TeV}).$ Such models predict a plethora of resonances, but the spectra and couplings of these particles are model dependent. A new discovery machine would be necessary to probe them. Unless we could directly probe Higgs scattering at energies above $f_\Pi,$ the smoking gun for these models would be deviations of the Higgs SM couplings. 

In particular, a relatively model-independent prediction is that the electroweak gauge boson mass terms become proportional to $\sin^2 \left(\frac{v + h}{f_\Pi}\right)$ where $v$ is the expectation value of the Higgs (no longer exactly equal to the value extracted from the electroweak fit $v_{\rm EW} \simeq 246$ GeV). This structure means that the couplings between the Higgs boson and the massive vector bosons are modified by a factor $\cos v/f_\Pi$ compared to their SM values, giving deviations of order $(v/f_\Pi)^2.$ Projections for the HL-LHC could constrain these couplings (interpreted as $\kappa_W, \kappa_Z$) to be within $1\%$ of the SM value \cite{deBlas:2019rxi}, giving a lower limit on $f_\Pi \gtrsim 2$ TeV, around the point at which the models start to become unnatural. A future lepton or hadron collider could then constrain them to be within $0.2\%$ of the SM values, or $f_\Pi \gtrsim 6$ TeV.

However, deviations of the vector boson couplings are not unique to composite Higgs models, and it would also be necessary to confirm any deviation by an observation of the \emph{triple-Higgs-coupling.} The deviations of this coupling from the SM value are harder to compute, and more model-dependent; but a calculation based on the potential  \cite{Contino:2010rs}
$$V(h) \simeq \alpha \cos\frac{v+h}{f_\Pi} - \beta \sin^2 \frac{v+h}{f_\Pi} 
$$
yields a deviation of the SM triple Higgs coupling $\kappa_3 = \cos v/f_\Pi$. The HL-LHC is only expected to give a sensitivity of order $50\%$ for this coupling, so a new high-energy collider is needed: the FCC-ee will not be able to access sufficient energy. A new hadron collider could reach a precision of a few percent, and therefore, in combination with the above measurements, would be able to confirm or refute whether compositeness is relevant for the electroweak scale. We hope to investigate this in more detail in future work. 

\section{The dark matter landscape}

\begin{figure}[h]
\includegraphics[width=16pc]{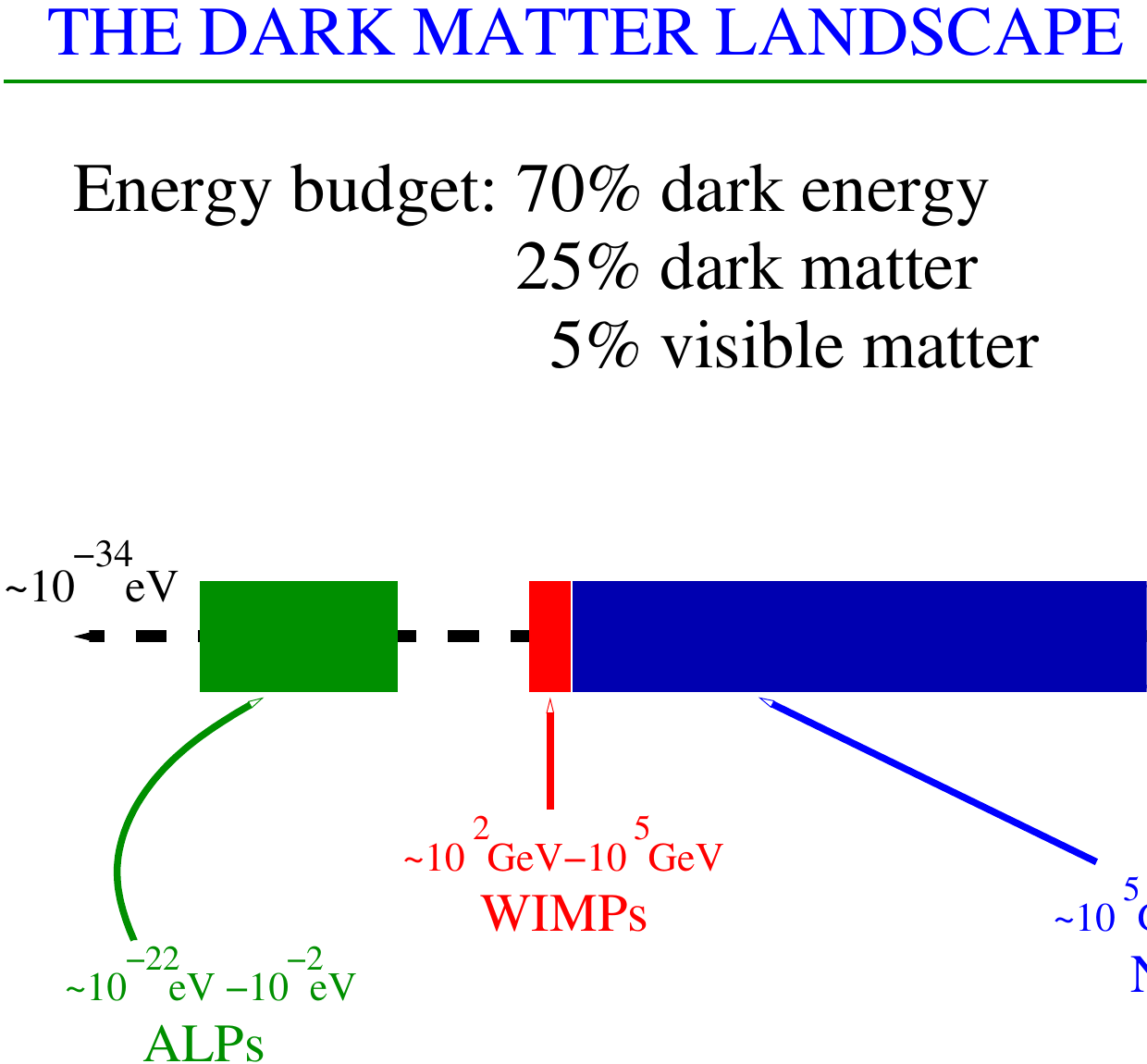}\hspace{2pc}%
\begin{minipage}[b]{10pc}
\caption{\label{darkmatterlandscape}%
\emph{
Dark matter candidates vary from ultra--light to ultra--massive, spanning more than eighty
orders of magnitude in mass scale. While the evidence for dark matter is compelling, 
searching for dark matter candidates is a vague proposition from an experimental 
point of view. 
}
}
\end{minipage}
\end{figure}

Among the contemporary puzzles of interest is the problem of dark matter. Astrophysical
and cosmological data indicate that the observed luminous matter in the universe accounts
for a fraction of the matter needed to explain the data, using the Standard Model
to parametrise the observations. This gives rise to the suggestion of numerous particle 
dark matter candidates that are sometimes motivated by other considerations, including: 
ultra-light dark matter; Weakly interacting Dark Matter; Self-interacting dark matter; 
Super-massive dark matter; axions; primordial black holes; ...
This bazaar of possibilities is depicted qualitatively in Figure 
\ref{darkmatterlandscape}.

String theory is a theory of gravity, which in our world couples to any form of
matter and energy. As a consequence, string models give rise to physics Beyond the
Standard Model (BSM) in various forms. One of the key predictions of string theory
is the existence of extra degrees of freedom, beyond those observed in the
Standard Model, that are required by the consistency of the theory. In some 
guise, these extra degrees of freedom take the form of extra dimensions, which, as we
illustrated in figure \ref{RSM}, leads to a compelling explanations for the
replication of three chiral generations in nature. The heterotic-string further
predicts the existence of additional gauge symmetries beyond the Standard Model.
The phenomenological fermionic $Z_2\times Z_2$ orbifolds give rise to a hidden
rank 8 gauge group. The Standard Model chiral generations that arise from the sectors
$b_1$, $b_2$ and $b_3$ are not charged under the rank 8 hidden sector gauge group.
The string models gives rise to additional $U(1)$ symmetries under
which the Standard Model states are charged. These additional $U(1)$ symmetries
are typically broken at a high energy scale. In some cases an extra $U(1)$ symmetry
may remain unbroken down to low scales \cite{frzprime}. 

A general consequence of the phenomenological string models is the existence of stable
string relics, Beyond the Standard Model (BSM) \cite{SSR}. These BSM states arise
due to the breaking of the non-Abelian $SO(10)$ GUT symmetry by Wilson lines.
The breaking of the GUT symmetry by Wilson lines produces states in the spectrum
that do not satisfy the quantisation conditions of the unbroken GUT group.
These states are dubbed as "Wilsonian matter states". One type of Wilsonian 
matter states that arise generically in the fermionic $Z_2\times Z_2$ 
orbifolds are states that carry fractional electric charge $\pm 1/2$ and 
are necessarily stable due to electric charge conservation.
Such states must appear in string models with the canonical 
GUT prediction $\sin^2\theta_W=3/8$. They appear in vector-like representations 
and can receive a heavy mass \cite{fc}. Curiously, there are also string
models in which the fractionally charged states do not arise as massless
states and appear only in the massive spectrum \cite{so64class}. 

In addition to the fractionally charged states, free fermion models in which the
$SO(10)$ symmetry is broken to $SU(3)\times SU(2)\times U(1)^2$ give rise to states
that carry fractional charges with respect to the $U(1)_{Z^\prime}\in SO(10)$ 
symmetry, but carry standard charges under the SM gauge group. Provided that the
extra $U(1)$ is broken by states that satisfy the $SO(10)$ quantisation rules
leaves a remnant discrete symmetry that forbids their decay into the SM states. 
These Wilsonian matter states can therefore provide viable dark matter candidates. 
Similar examples arise in string model of ref. \cite{frzprime} that allows for 
the $U(1)$ in $E_6\rightarrow SO(10)\times U(1)_A$  to remain unbroken down to low 
scales. This $U(1)$ symmetry is anomalous in many strings constructions
\cite{auone}. The breaking of $E_6$ in the string models 
results in the projection from the massless spectrum of some components
from the chiral $E_6$ representations. The massless spectrum forms incomplete 
$E_6$ representations rendering $U(1)_A$ anomalous. In \cite{frzprime} self--duality 
under spinor--vector duality \cite{svd} was used to construct a model with
broken $E_6$ and complete $E_6$ multiplets and in this case $U(1)_A$
is anomaly free and can remain
unbroken down to lower scales. 

The string derived $Z^\prime$ model provides an example to illustrate the
ambiguity in pinning down the particle properties of dark matter candidates. 
The unbroken $SO(10)$ subgroup at the string level in the model
is the Pati--Salam subgroup $SO(6)\times SO(4)$, which is broken down by the VEV of 
heavy Higgs states in the model. The remaining unbroken combination is given by:
\begin{equation}
    U(1)_{Z^\prime}=\frac{1}{5}\left(U(1)_C - U(1)_L  \right) -U(1)_A
    \label{u1zprime}
\end{equation}
where $(U(1)_C-U(1)_L)\in SO(10)$ is orthogonal to the weak hypercharge. 
The spectrum below the Pati-Salam breaking scale is given in table 
\ref{lowzpspectrum}. 

\begin{table}[h]
\noindent 
{\small
\begin{center}
{
\begin{tabular}{|l|cc|c|c|c|}
\hline
Field &$\hphantom{\times}SU(3)_C$&$\times SU(2)_L $
&${U(1)}_{Y}$&${U(1)}_{Z^\prime}$  \\
\hline
$Q_L^i$&    $3$       &  $2$ &  $+\frac{1}{6}$   & $-\frac{2}{5}$   ~~  \\
$u_L^i$&    ${\bar3}$ &  $1$ &  $-\frac{2}{3}$   & $-\frac{2}{5}$   ~~  \\
$d_L^i$&    ${\bar3}$ &  $1$ &  $+\frac{1}{3}$   & $-\frac{4}{5}$  ~~  \\
$e_L^i$&    $1$       &  $1$ &  $+1          $   & $-\frac{2}{5}$  ~~  \\
$L_L^i$&    $1$       &  $2$ &  $-\frac{1}{2}$   & $-\frac{4}{5}$  ~~  \\
%
\hline
$D^i$       & $3$     & $1$ & $-\frac{1}{3}$     & $+\frac{4}{5}$  ~~    \\
${\bar D}^i$& ${\bar3}$ & $1$ &  $+\frac{1}{3}$  &   $+\frac{6}{5}$  ~~    \\
$H^i$       & $1$       & $2$ &  $-\frac{1}{2}$   &  $+\frac{6}{5}$ ~~    \\
${\bar H}^i$& $1$       & $2$ &  $+\frac{1}{2}$   &   $+\frac{4}{5}$   ~~  \\
\hline
$S^i$       & $1$       & $1$ &  ~~$0$  &  $-2$       ~~   \\
\hline
\hline
$h$         & $1$       & $2$ &  $-\frac{1}{2}$  &  $-\frac{4}{5}$  ~~    \\
${\bar h}$  & $1$       & $2$ &  $+\frac{1}{2}$  &  $+\frac{4}{5}$  ~~    \\
%
%
%
\hline
$\phi$       & $1$       & $1$ &  ~~$0$         & $-1$     ~~   \\
$\bar\phi$       & $1$       & $1$ &  ~~$0$     & $+1$     ~~   \\
\hline
\hline
$\zeta^i$       & $1$       & $1$ &  ~~$0$  &  ~~$0$       ~~   \\
\hline
\end{tabular}}
\end{center}
}
\caption{\label{lowzpspectrum}
\it
Spectrum and
$SU(3)_C\times SU(2)_L\times U(1)_{Y}\times U(1)_{{Z}^\prime}$ 
quantum numbers, with $i=1,2,3$ for the three light 
generations. The charges are displayed in the 
normalisation used in free fermionic heterotic-string models. }
\end{table}

From Table \ref{lowzpspectrum} we note the states ${\phi}~{\rm and}~{\bar\phi}$ versus the
$S^i$ states, which are the $SO(10)$ singlets in the $27$ representation of $E_6$. 
The $U(1)_{Z^\prime}$ charge of the $S^i$ singlets is $-2$, whereas that of
${\phi}~{\rm and}~{\bar\phi}$ is $\pm1$, 
which are Wilsonian matter states and candidates
for dark matter \cite{wdm} provided that $U(1)_{Z^\prime}$ is broken by a VEV of $S^i$ that carries the standard $E_6$ charge. The 
${\phi}~{\rm and}~{\bar\phi}$ states arise because of the breaking of $E_6$
by Wilson lines in the string models.

The Wilsonian matter states in the string derived $Z^\prime$
model provide an example to illustrate the vagueness of the 
dark matter proposition from an experimental point of view 
\cite{wdm}. We list below several cases to consider: 
\begin{enumerate}
\item $M\gg M_{Z^\prime}$ without inflation~~~~~~~~~{$\Rightarrow M\le 10^{5}~\hbox{GeV}$}
\item { $M\gg M_{Z^\prime}$ with inflation and $T_R > M_{Z^\prime}$~~~~~
{$\Rightarrow M > T_R\left[25+{{1}\over{2}}\ln\left({{M}\over{T_R}}\right)\right].$}}
\item  { $M\ll M_{Z^\prime}$ without inflation ~~~~~~~
{$\Rightarrow M<3~\hbox{keV}$}}
\item { $M\ll M_{Z^\prime} ~~\hbox{with inflation}~~
M~~
\begin{cases}
>~T_R\left[25+{{1}\over{2}}\ln\left({{M^5}\over{M_{Z^\prime}^4T_R}}\right)\right]~,
                 &   T_R < M \cr%
<~{{M^4_{Z^\prime}\over{T_R^3}}6.9\times 10^{-25}%
\left({{g_*}\over{200}}\right)^{1.5}%
{{1}\over{N_{Z^\prime} g^2_{\rm eff}}}},& T_R > M
\end{cases}
$}
\end{enumerate}

where $M$ is the dark matter mass and $M_{Z^\prime}$ is the mass of the $Z^\prime$
vector boson. The cases to consider include the possibility of inflation and
whether $M< M_{Z^\prime}$ or $M> M_{Z^\prime}$. In the first case $M\gg M_{Z^\prime}$
without inflation. In this case the Wilsonian dark matter is in thermal equilibrium 
in the early universe and is constrain to be below $10^{5}$GeV. This case is
similar to neutralino dark matter. In the second case $M>M_{Z^\prime}$ but
the existence of inflation is assumed. In this case the Wilsonian dark matter
is diluted by inflation and is reproduced by reheating to the required 
phenomenological abundance. This gives a relation between the Wilsonian 
dark matter mass and the reheating temperature and the Wilsonian dark
matter can be superheavy. In the third case, $M<M_{Z^\prime}$ without 
inflation. In this case, the Wilsonian dark matter candidate behaves 
like warm dark matter candidate and is constrained to be below 3 keV. Finally,
in the case $M<M_{Z^\prime}$ with inflation, there is again a complicated relation
between $M,~M_{Z^\prime}$ and the reheating temperature $T_R$. This demonstrates
that even in this quite concrete case, there is a vast range of possibilities 
with different experimental strategies to try to detect the Wilsonian 
dark matter candidate. In addition to the Wilsonian dark matter candidates, 
other dark matter candidates that may arise from the string models include: 
hidden sector glueballs \cite{fp} and axion-like candidates.
Our conclusion is that, while continued experimental 
searches for dark matter are certainly worthwhile, the 
particle characteristics of the dark matter candidates are not pinned down 
from an experimental point of view.

\section{Future Collider Facilities}

We argued that the study of the Higgs particle presents a well defined experimental program. 
Furthermore, from a string phenomenology perspective, the experimental
resolution of the nature of the Higgs boson is crucial to indicate
whether the construction of phenomenological string models that has been
pursued over the past 30+ years, which rests on the assumption that
the Standard Model Higgs boson is a fundamental scalar particle, is relevant
in nature. The experimental priority therefore should be the experimental
studies of the Higgs sector of the Standard Model and beyond. 

The charting of the future of the experimental particle physics
program has been under communal scrutiny over the past few years
\cite{Butler:2023eah, Gourlay:2022odf}. 
We have also contributed to this discourse via participation 
in community meetings as well as via publications dedicated to this
question, where we used the string derived $Z^\prime$ model as a 
bench mark model to illustrate some of the arguments. We will recap these
arguments here briefly and further discussion can be found in our
earlier publications \cite{fgMc, thequest}. 

The discussions revolves around several key points that can be divided
into technological and social.
The first social issue is regional, 
with the main proponents being China; the European Union; and the United States, 
which possess in principle the know how to develop the desired facilities. 
Additional players with relevant expertise are Japan and Russia that are at 
a less committed stage of development for future collider facilities 
A key technological issue is whether to prioritise the precision
frontier versus the energy frontier, where the first means
prioritising lepton colliders whereas the second entails prioritising hadron colliders. 
Another key aspect revolves on the development of accelerator
physics technology. This takes several forms. One is the development
new superconducting alloys that can sustain stable magnetic fields of the
order of 16 Tesla, whereas the superconducting alloys that are used at the
LHC operates at 8.3 Tesla and might be able to sustain magnetic fields of 
the order of 10 Tesla. Another aspect of magnet technology is the
development of HTC magnets that can operate at liquid nitrogen temperature,
which will dramatically cheapen the operation of the superconducting magnets.
Another challenging accelerator physics technology is the development of a
muon collider. As with all developments of new technology they make the
timeline for the project open-handed.
All these accelerator physics projects are fascinating in their own
rights and ought to be pursued. They will bring many applications 
and benefits outside the realm of particle physics.
The main question is that of the priorities
of the experimental particle physics program. 

In Europe the LHC and High--Luminosity LHC will operate at 14 TeV CoM
energy until the early 2040s. It will do fascinating Bread \& Butter 
(B\&B) physics. It will improve the measurements of key Standard
Model parameters and will constrain many BSM scenarios, with substantial
capacity of discovery in case new physics exists within its energy reach. 
During that period, it is expected that developments at CERN will follow the LEP playbook. 
A 90--100km tunnel will be dug for the Future Circular Collider (FCC).
The first phase of the project will be a high luminosity $e^+e^-$ collider
that will operate initially at 250 GeV and at later stages at 350 GeV.
The FCC-ee
will perform precision measurements of the Higgs and top parameters
in the kinematically accessible energy range, {\it e.g.} it will not 
be able to measure directly the $t{\bar t}h$ coupling. Following the
leptonic phase, an hadronic FCC--$hh$ collider 
at 90--100 TeV will be operational from the early 2070s.
The FCC--$hh$ will be have more substantial capacity to
study BSM physics and to measure SM parameters that are 
beyond the reach of the FCC--ee. 

The Chinese program with CEPC and SPPC mirrors the European program with 
an initial circular lepton collider followed by a hadron collider, with the caveat
that the CEPC may become operational 5 years prior to FCC-ee. 
This may force CERN to revise its plans, which is an on going process.

An in-depth exercise to examine its future particle physics program
was conducted in the US, and cumulated in the Snowmass 2021 report
\cite{Butler:2023eah}. The focus on neutrino physics will
continue in the near term and the long term recommendation of the 
Particle Physics Project Prioritization Panel (P5) is to build a
muon collider. Building a muon collider presents major
technological challenges. Given the experience of the MICE
experiment \cite{Gregoire:2003nh, MICE:2019jkl}, which demonstrated the feasibility 
of the muon beam phase space cooling, it is fair to assess that
a muon collider may become operational as a physics machine
in the future. 

Another major contender on the planning boards is the International Linear Collider
(ILC), which is a linear $e^+e^-$ collider. The main selling point of this
option is that a linear lepton collider can reach higher CoM energies, 
possibly up to 1TeV. This will enable probing the Higgs sector at higher energies
compared to the FCC--$e^+e^-$ and more precisely relative to a circular
hadron collider. The drawback is the reduced luminosity compared to a
a circular lepton collider. 
Similar 
to the FCC--ee, the ILC will operate at different stages, {\it i.e.}
at 90 GeV, 250 GeV, 350 GeV, 500 GeV, and 1 TeV, and the time interval between the
different phases means that this will be a prolonged project over many
decades. 

Two additional considerations are worthy of mention. 
The first is the notion that a new collider experiment must guarantee a discovery of new physics.
In our view, the main deliverable of a Future Collider Facility is
to improve the measurement of the Standard Model parameters and to
measure the parameters of the Standard Model that are not
accessible in current facilities, {\it e.g.} the triple Higgs coupling and
the $t{\bar t}h$ coupling, given that there is no clear indication what the new
physics should be. 
Discovering new physics is obviously to be hoped for, and indeed it is anticipated that new physics associated with electroweak symmetry breaking should exist, along with the argument that we will never uncover the answers to the most fundamental questions if we do not test our theories in previously unexplored regions. Hence, such discoveries are necessarily unexpected.

The second consideration is the timeline of new projects. Particle physics
experiments have become a prolonged affair, and might soon compete on
that front with inter-stellar travel. While traveling to other solar
systems is still in the realm of fantasy, planning of particle physics
experiments is real. The sustainability, of such long term programs,
and in particular the training and maintaining of Early Career Researchers
and their longer term interest in the field is a concern.

\begin{figure} 
\begin{center}
\includegraphics[width=7.cm]{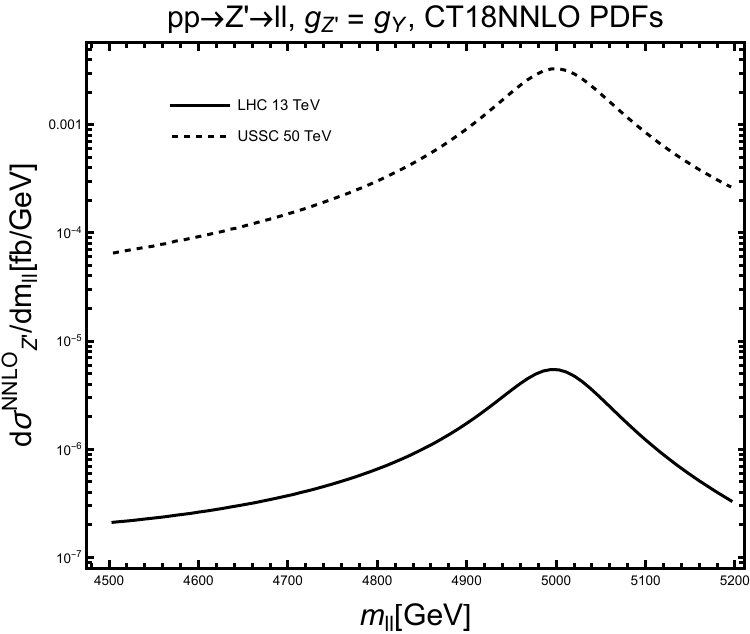}
\caption{$Z^\prime$ production and leptonic decay at 14TeV and 50TeV
(figure taken from arXiv:2309.15707 \cite{fgMc}) }
\label{zpatpp}
\end{center}
\end{figure}

An alternative route for Future Collider Facilities (FCF) is a hadronic collider
that utilises established magnet technology. In \cite{fgMc, thequest} we dubbed this
proposal as Upgraded Superconducting Super Collider (USSC), or alternatively 
as FCC--LHC. The basic parameters of the USSC/FCC--LHC are close to the defunct
SSC that was supposed to have an 87.1km circular ring with 
6.3Tesla magnets and to operate at 40TeV CoM energy. Our proposal
is that with established but improved magnet technology, one can envision
a hadron collider experiment with 50--60TeV CoM energy in a 91km FCC-like tunnel. 
The USSC/FCC--LHC will be able to do Bread \& Butter Standard Model
measurements, including the triple Higgs and the $t{\bar t}h$ couplings
and will have substantial capacity to discover physics Beyond
the Standard Model, if it exists. In \cite{fgMc} we showed that the
cross sections at the USSC/FCC--LHC with 50TeV CoM energy increases by three
orders of magnitude compared to the LHC. 
Figure (\ref{zpatpp}) demonstrates this increase for the $Z^\prime$
production and leptonic decay at 14TeV and 50TeV.  

\begin{figure}[t]\centering
\includegraphics[width=0.45\textwidth]{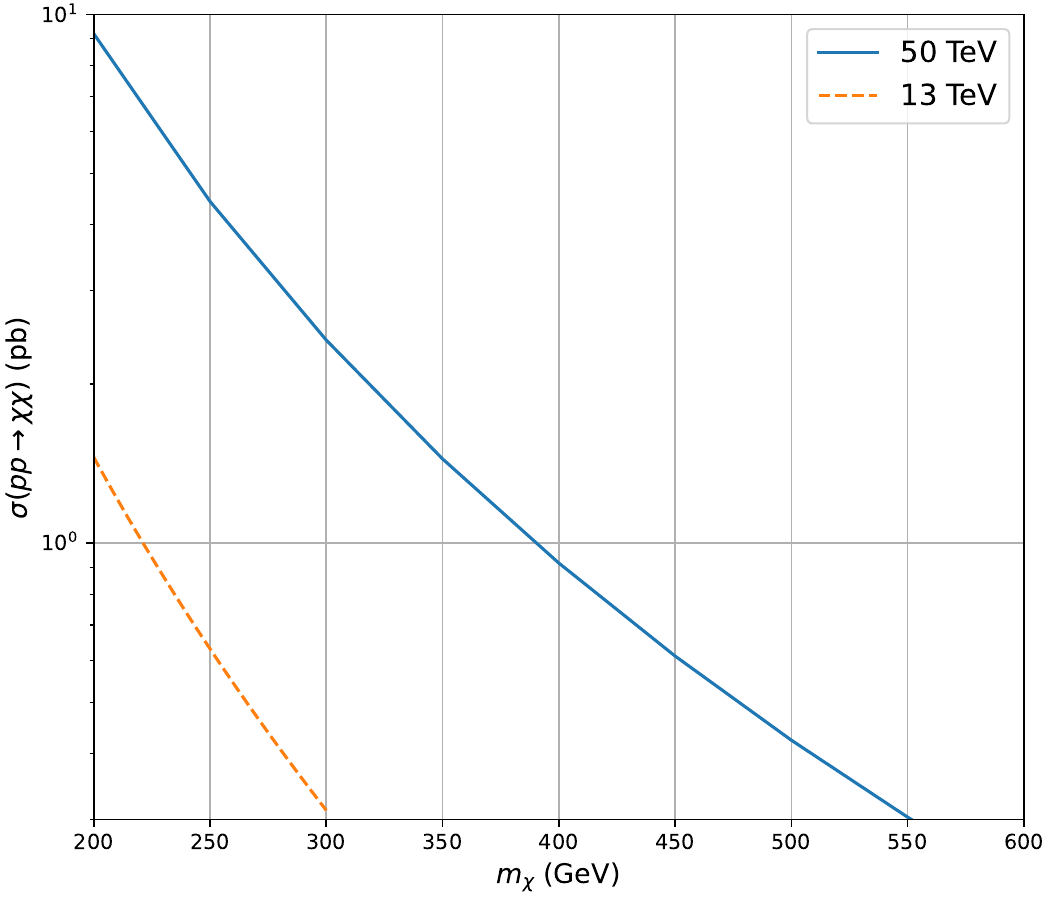}\includegraphics[width=0.45\textwidth]{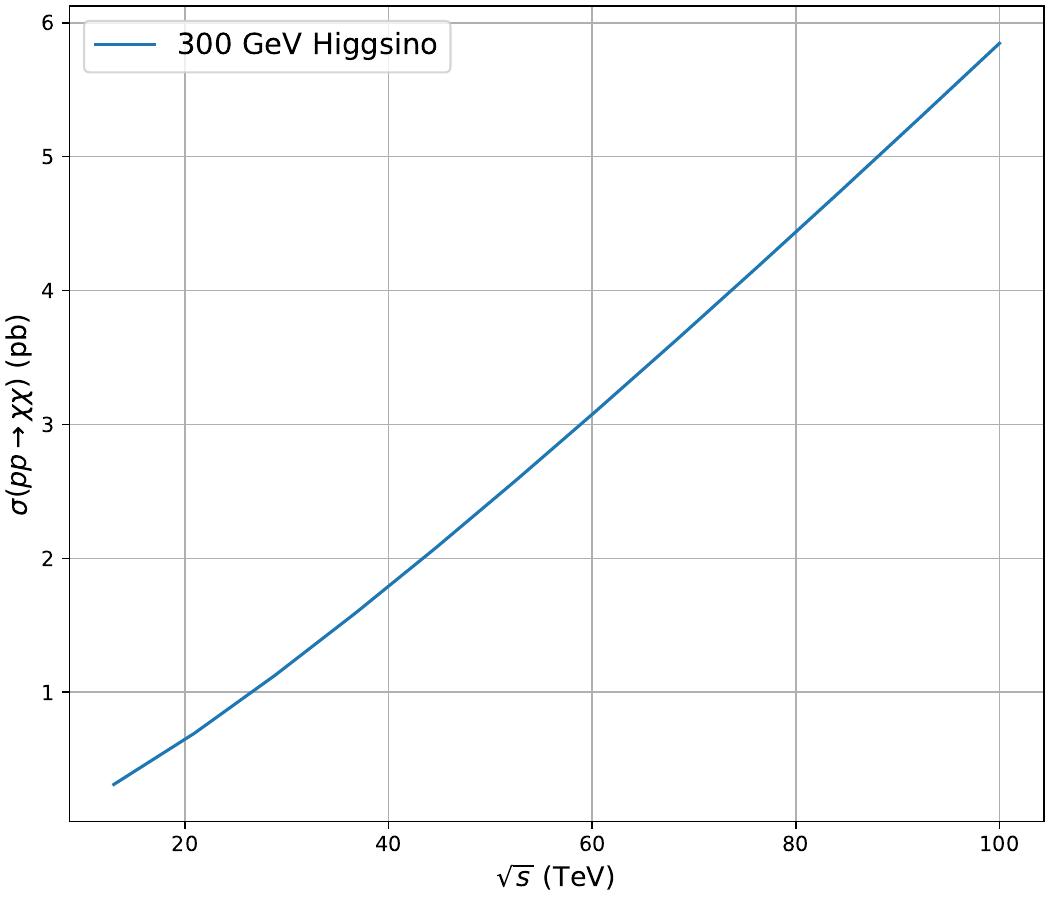}
\caption{\label{FIG:HinoXS}NLO cross-sections for degenerate higgsino pair production (left) varying the Higgsino mass and (right) varying collider centre-of-mass energy.}
\end{figure}

Indeed, a future hadron collider would produce electroweak-charged particles abundantly, with a significant enhancement compared to the LHC -- and at a mass scale inaccessible to planned future lepton colliders. Such particles could be associated with either dark matter or electroweak symmetry breaking.  As can be seen in figure \ref{FIG:HinoXS} the cross-section for (degenerate) Higgsinos -- one of the hardest and most interesting such cases, because they (or similar particles) may be responsible for \emph{four} anomalies in the Run 2 data \cite{Agin:2023yoq,Agin:2024yfs} -- is dramatically enhanced between 13 and 50 TeV centre-of-mass energy.

Given that the SSC price tag
was \$6B and allowing for inflation since its cancellation puts an estimate
of \$25B on the cost of the USSC/FCC--LHC. Given that the site selection of the
Original SSC (OSSC) was made in October 1999 and that it was supposed to start
operations in the 1996--1999 period, the USSC/FCC--LHC can materialise in 10--15
years from decision, {\it i.e.} it can start delivering physics results from the
late 2030s.

The main regional players in experimental particle physics (China; Europe; Japan; United States; Russia)
face an abundance of riches that clouds the decision on the course of action. Additionally, the notion
that the success of a collider experiment rests on the discovery of new physics hinders the
commitment and investment in a project of such magnitude. The discovery of new physics is
not guaranteed. In regions where the basic technological know--how already exists, it makes sense,
given the projected cost of a new collider experiment, to prioritise technology development, 
{\it e.g.} of new superconducting alloys that enable the construction of magnets with
stronger magnetic fields. The new magnet technology will have many applications in the wider economy.

This state of affairs gives an opportunity for new players that do not currently
possess the required accelerator physics know-how and infrastructure to enter the field.
Given that the required magnet technology is off the shelf technology,
we proposed in \cite{fgMc, thequest} that the USSC/FCC--LHC can be pursued as a
Middle East project at the SESAME site in Jordan and funded by Saudi-Arabia and
other regional countries. The potential benefit in terms of international prestige
and technological development is substantial.
The construction of the USSC/FCC--LHC and successful delivery
on the specified design parameters would be an enormous success. Following the CERN experience of
providing a platform for cooperation, the USSC/FCC--LHC can serve as a Project for Peace, 
promoting curiosity driven collaboration between neighbouring and far away countries. The discovery
of new physics at the USSC/FCC--LHC will not only transform the field of particle physics,
but will change the "Face of Arrakis" promoting the regional sponsors into the
world leaders in the quest of humanity to understand the universe in which we live and the rules
that govern it.

\section*{Acknowledgements}

MDG is supported in part by Grant ANR-21-CE31-0013, Project DMwithLLPatLHC, from the \emph{Agence Nationale de la Recherche} (ANR), France. He acknowledges helpful discussions with Werner Porod, Giacomo Cacciapaglia and Benjamin Fuks. We thank Karim Benakli for related conversations.  
This work was supported by a Royal Society Exchange grant IES/R1/221199, ``Phenomenological studies of string vacua". The work of M.G., and J.D. has been supported by the National Science Foundation under Grant No. PHY 2412071.

\end{document}